\documentclass[reprint,preprintnumbers, amsmath,amssymb, aps]{revtex4-2}

\usepackage{graphicx}% Include figure files
\usepackage{dcolumn}% Align table columns on decimal point
\usepackage{bm}% bold math
\usepackage{changes}
\usepackage{soul}
\usepackage{amsmath}
\usepackage{color}

\begin{document}

\title{Omnidirectional wave impedance matching and Goos–H\"anchen shifts in non-Hermitian balanced positive–negative index metamaterials}

\author{Jeng Yi Lee$^{1}$, Chia-Heng Sun$^{2}$, Ke-Hsiang Huang$^{1}$ and Pai-Yen Chen$^{2}$}

\affiliation{$^{1}$Department of Opto-Electronic Engineering, National Dong Hwa University, Hualien 974301, Taiwan,\\
$^{2}$Department of Electrical and Computer Engineering, University of Illinois at Chicago, Illinois 60607, USA
}

\begin{abstract}
Wave reflection  by non-Hermitian antisymmetric parity-time (APT) photonics made of a balanced positive index material (PIM) and negative index material (NIM) is investigated.
As one class of them satisfies a unitary relation, wave impedance matching (WIM) can be achieved not only for polarization independence but also for omnidirection.
Remarkably, the result is regardless of system sizes, material parameters, and polarization.
As gain or loss are involved, any non-Hermitian systems are unable to support WIM; instead, there occurs a reflection dip, i.e., a minimum reflectance, at some system configurations, a certain incident angle, and polarization dependence.
  Our finding is in contrast to the parity time (PT) symmetric paradigms, in which the well known exceptional point can be recognized as a WIM.
Additionally, in vicinity of a reflection dip, it is accompanied by a jump of reflection phase, offering an opportunity to observe a significant Goos-H\"anchen (GH) beam shift in APT  systems.
We provides associated systems and numerically verify the GH shifts.
 Meanwhile, we derive a closed-form expression for a reflection dip in terms of incident angle, operating wavelengths, and material parameters.
To imitate the electromagnetic responses of APT photonics for their angular dispersion, we discuss a dielectric heterostructure consisting of spatially symmetric multilayer slabs  so as to have the same reflection and transmission for each APT component over a specific angular spread.
\end{abstract}

\maketitle

 \section{Introduction}
Wave reflection at an interface of two adjacent different media is driven by wave impedance mismatch.
Formation of omnidirectional wave impedance matching (OWIM) i.e, wave impedance matching for all incident angles, is challenge, that relies on specific designs or unusual electromagnetic materials \cite{OMN1,OMN2,OMN3,OMN4,OMN5,superlens1}.
For instance, by prescribing wave-impedance boundary conditions at each interfaces and exploiting an ansatz for the spatial distribution of material parameters, spatiotemporal modulated media are derived from reduced wave equations, thereby enabling OWIM \cite{OMN1,OMN2}. 
Engineering photonic crystals with a shifted elliptically spatial dispersion, 
such photonic crystals exhibit  not only  transparent media but also aberration-free virtual images \cite{OMN5}.
As a single planar slab in free space exhibits OWIM, it can be achieved by simultaneously relative permittivity $\epsilon=-1$ and relative permeability 
$\mu=-1$ \cite{superlens1,superlens2}.
Hence, the refracted ray is reversed and the evanescent waves are amplified, allowing the planar slab to function as a lens and surpassing the diffraction limit \cite{superlens1}.
Interestingly, the antiparallel orientation between the wave vector and the Poynting vector is not an exclusive signature of negative-index materials (NIMs), while it can also occur in certain lossy dielectric–magnetic media \cite{condition1,condition2}.

Evidently, extending the standard Hermitian systems into non-Hermitian regimes is foresight, as they can reveal extraordinary wave phenomena and thereby inspire novel wave-based designs \cite{non1,non2,non3}.
A representative manifestation is parity–time (PT) symmetric systems made of a balanced gain-loss, moving forward  development of novel devices with unique
functionalities, such as negative refraction \cite{nrPT}, stably high efficient power transfer \cite{PTpowertransfer1,PTpowertransfer2,PTpowertransfer3}, ultra-sensitive sensors \cite{PTsensor1,PTsensor2,PTsensor3}, cryptographic key generation \cite{PTPUF1,PTPUF2,PTPUF3} to name a few.
Significantly, one unique feature brought from non-Hermitian systems is an exceptional point, which is characterized by the coalescence of eigenvalues and their corresponding eigenstates \cite{EP1}, in contrast to the diabolic points in Hermitian systems.
In PT-symmetric systems, the exceptional point characterized by $R=0$ and $T=1$ is essentially a wave impedance matching \cite{PT2}, which is polarization sensitivity as well as incidence dependence.
Regarding reflection,  the reflection phase in the vicinity of the exceptional point undergoes a $\pi$ jump, which can be exploited to enhance the transverse spin Hall shift \cite{PT3}. 
Moreover, surface polaritons can arise at a non-Hermitian interface in a stable and undissipative manner, accompanied by anomalous energy flux, spin-orbit angular momentum locking, and abrupt reflection-phase variations \cite{PT4,PT5}.
A balanced gain-loss metasurface can have one-sided wave impedance matching, thereby providing unidirectional reflectionless \cite{PT1}.  
In complex PT-symmetric crystals, reflection in stop band can be amplified \cite{PTcrytal1,PTcrytal2}.
More appealingly, that becomes significantly enhanced in broken PT-symmetric phases.

 \begin{figure*}[ht]
 \centering
\includegraphics[width=1\textwidth]{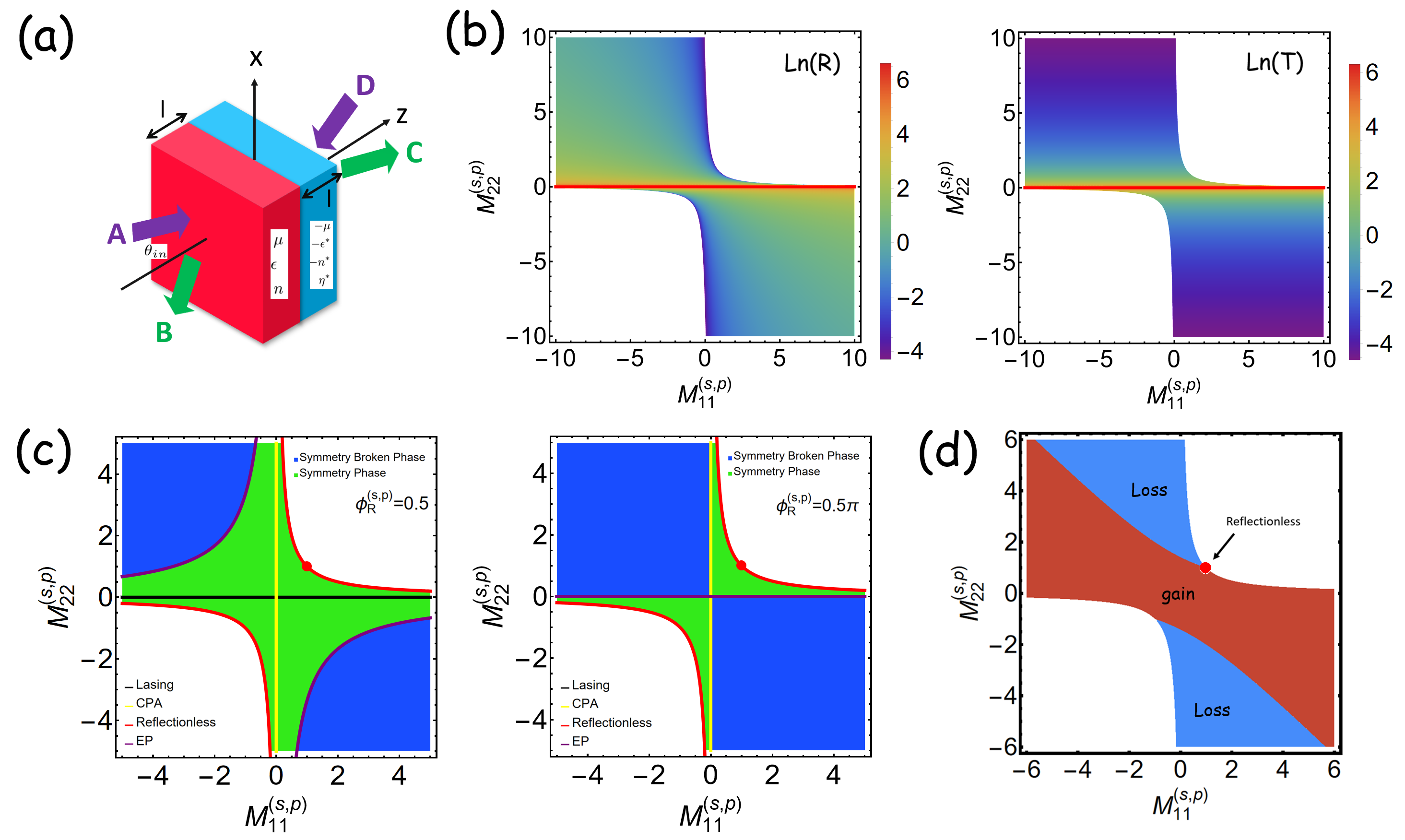}
\caption{(a) Schematic of  a bilayer configuration with APT condition.  By the parametrization, we depict the logarithm
of reflectance $ln(R)$ and the logarithm of transmittance $ln(T)$ in (b). (c) By taking  $\phi^{(s,p)}_R=0.5$ rad and $\phi^{(s,p)}_R=0.5\pi$ rad, we display the regions with symmetry phase marked by green, EP by purple lines, and symmetry broken phase by blue.
Here, the yellow line, the black line, and the red lines are denoted as coherent perfect absorption, lasing, and reflectionless, respectively. 
By considering energy conservation as well as the parametrization, we illustrate light blue region for lossy bilayer APT systems, dark red one for amplifying ones, and a green dot for lossless ones in the parametric space of (d). }
\end{figure*} 

On the other hand, another closely related non-Hermitian system involving NIMs satisfies $n(-x)=-n^{*}(x)$, known as an antisymmetric parity–time (APT) system, which was first explored in \cite{APT1}.
APT systems enable a flat transmission band \cite{APT1}, a continuous lasing and coherent perfect absorption \cite{APT2,APT3}, approaching the quantum limit of sensing and squeezing \cite{APT4}, constant refraction in coupled waveguides \cite{APT5,APT6}, energy-difference conserving
in electric circuits \cite{APT7}, and superior sensing as well as strong robust against error by APT-based electric elements \cite{APT8}.
Interestingly, although  the concept based on APT   has found successful exploration and applications in  many physical disciplines, such as acoustics,  electrical circuits, quantum entanglement to name a few \cite{APT1,APT2,APT3,APT4,APT5,APT6,APT7,APT8,APT9}, a complete understanding of reflection in APT  photonics is still lacking.

In this work, we investigate the electromagnetic wave reflection by a bilayer slab configuration having  APT condition, where their refractive index satisfies $n(-x)=-n^{*}(x)$.
To elucidate their limits, we first provide a parametric space, where is derived from a consideration of APT symmetry as well as reciprocity principle.
We observe that when one class of APT systems satisfies a unitary relation, the corresponding systems can enable OWIM. 
Interestingly, this finding is irrespective of incident polarization, refractive indices, and system sizes.
To further understand the behind mechanism, we provide wave impedance analysis. 
We state that as  loss or gain is involved,  the OWIM condition is broken.
Instead, there occurs a reflection dip depending on specific system configurations, operating wavelengths, incident polarization, and incident angles.
We derive an analytical expression to indicate their formation.
Additionally, a jump of reflection phase by $\pi$ is observed in vicinity of this dip, offering an opportunity to observe an enhanced Goos–H\"anchen beam shift.
Accordingly, we numerically simulate the phenomena by employing a Gaussian beam upon the corresponding systems. 
Interestingly, natural materials exhibiting simultaneous negative permeability and negative permittivity had found in Dirac semimetals such as  Cd$_{3}$As$_2$ at gigahertz frequencies \cite{effective1}.
 However,  APT systems require strict material parameters, which become an obstacle to being realized.
Therefore, we discuss an all-dielectric heterostructure with a spatial symmetry to imitate the electromagnetic responses of each APT slab component in their angular dispersion of reflection.
Specifically, the mimicking heterostructure can nearly generate the same electromagnetic transmission and reflection fields of APT systems over a finite angular range. 
We remark that similar architectures have previously been explored under normal incidence, as  in Ref. \cite{APT3}.
We believe that our work can promote potential applications, including optical sensing, subwavelength imaging, precise metrology, spin-based devices, radar stealth, and invisibility cloaking to name a few.

\section{Theoretical formalism}
\subsection{System Configuration, Transfer Matrix Parametrization, and Parametric Space}
We consider a bilayer slab system having APT condition, namely $n(-x)=-n^{*}(x)$, embedded in free space, as in Fig. 1(a).
Following Ref.\cite{APT1}, the relative permeability is accordingly with $\mu(-x)=-\mu(x)$, while the relative permittivity satisfies $\epsilon(-x)=-\epsilon^{*}(x)$. 
Within a bilayer configuration, the APT systems have purely lossless, purely lossy, or purely gain.
Without loss of any generality, we let the left slab made of a positive-index material (PIM) and the right one made of a negative-index material (NIM), while the thickness of both slabs is accordingly set to be $l$.

We discuss a linearly polarized electromagnetic plane wave obliquely incident upon ours system, where the incident angle is denoted by $\theta_{in}$.
Then, we let the plane of incidence be embedded in the x-z plane.
The incident polarization can be  either  TE polarization (\textbf{s}, $\vec{E}\parallel \hat{y}$) or TM polarization (\textbf{p}, $\vec{H}\parallel \hat{y}$).
The complex amplitudes of the incident  incoming plane waves are represented by $A^{(s,p)}$ and $D^{(s,p)}$, while those of the outgoing scattering waves are by 
$B^{(s,p)}$ and $C^{(s,p)}$.
Here the superscripts by $(s,p)$ are exploited to distinguish two different polarization incidences.
Throughout this work, we choose the time harmonic of these waves be $e^{-i\omega t}$, where $\omega$ is angular frequency. 

Next, we analyze the transfer matrix with APT, denoted by $M^{(s,p)}$, while it reads $[C^{(s,p)},D^{(s,p)}]^T=M^{(s,p)}[A^{(s,p)},B^{(s,p)}]^T$.
Here, $T$ represents the transpose operator.
On the other hand, unlike PT-symmetric paradigms, APT systems lack any explicit physical symmetry operators.
Instead, one can directly observe that the APT-symmetric transfer matrix satisfies  $M^{(s,p)*}\sigma M^{(s,p)}=\sigma$. 
Here $\sigma$ is parity operator.
We provide the detailed derivation in Appendix A.

Benefiting from this relation, we can thus introduce the parametrization for $M^{(s,p)}$.
Moreover, since the APT system is embedded in a symmetric environment, the reciprocity principle holds.
We can obtain that the parametrization for arbitrary APT-symmetric transfer matrix can be in terms of three independent real variables, namely,  $M^{(s,p)}_{11}$, $M^{(s,p)}_{22}$, and $\phi^{(s,p)}_R$ \cite{APT3}.
Here $\phi^{(s,p)}_R$ is denoted as the reflection phase for right incidence.
Strikingly, we find $M^{(s,p)}_{11}M^{(s,p)}_{22}\leq 1$.
Meanwhile, the transmission coefficient $t$, reflection coefficient at left side $r_L$, and reflection coefficient at right side $r_R$ can be generally expressed in terms of parametrization,
\begin{equation}
\begin{split}
\begin{cases}
t^{(s,p)}&=\frac{1}{M^{(s,p)}_{22}}\\
r^{(s,p)}_L&=\frac{\sqrt{1-M^{(s,p)}_{11}M^{(s,p)}_{22}}}{M_{22}}e^{-i\phi^{(s,p)}_R}\\
r^{(s,p)}_R&=\frac{\sqrt{1-M^{(s,p)}_{11}M^{(s,p)}_{22}}}{M_{22}}e^{i\phi^{(s,p)}_R}
\end{cases}
\end{split}
\end{equation}
\cite{APT3}.
Therefore, we can state that the phase of $t$ can be only in-phase or out-of-phase, while $r^{(s,p)}_L=r^{(s,p)*}_R$ represents that their magnitudes are identical and their phases form a pair with opposite signs.

To understand the limits of reflectance $\vert r\vert^2$ and transmittance $\vert t\vert^2$ for any APT systems, with the parametrization, we present a parametric space in Fig. 1 (b).
The result is independent of specific system configurations, frequency, and material parameters.
We mark that various colors correspond to different values of their logarithm, whereas the white region represents parameters that are not allowed by APT condition.
Most notably, the reflectionless appears at the boundary between the color region and the white one, where satisfies
 $M^{(s,p)}_{11}M^{(s,p)}_{22}=1$.
We also discover that any APT systems with reflectionless can enable different levels of transmittance.
In contrast, PT-symmetric systems with reflectionless would be accompanied by a unit transmittance, since they obey the conservation relation \cite{PTphase2}.
In addition, we observe that both  $R^{(s,p)}$ and $T^{(s,p)}$ can diverge to infinity, as marked by a red line of Fig. 1 (b), corresponding to lasing, which had explored in Refs. \cite{APT1,APT2,APT3}.
Moreover, APT systems also support coherent perfect absorption,
which two counterpropagating incident waves with appropriate amplitudes and phases are completely absorbed, already stated and explored in Refs. \cite{APT1,APT2,APT3}.

\subsection{Scattering matrix}
By analysis of the scattering matrix, i.e., $S$ matrix, PT-symmetric architectures can have a symmetry phase with norm power preserving or a symmetry broken phase having a net power amplification and dissipation \cite{PTphase1}.
In between, there is an exceptional point (EP), charactered by the coalescence of both scattering eigenvalues and their scattering eigenvectors \cite{EP1}.
However, there have two different expressions for $S$ matrix, leading to different scattering phase explanations \cite{PTphase2,PTphase1}.

Here, the $S$ matrix we employ  is written as $\begin{bmatrix}B^{(s,p)}\\C^{(s,p)}\end{bmatrix}=S^{(s,p)}\begin{bmatrix}A^{(s,p)}\\D^{(s,p)}\end{bmatrix}$, where $S^{(s,p)}=\begin{bmatrix} r^{(s,p)}_L & t^{(s,p)} \\ t^{(s,p)} & r^{(s,p)}_R \end{bmatrix}$ is a symmetric matrix and  possesses pseudo-Hermitian \cite{ps,APT1}.
With the parametrization, the corresponding scattering eigenvalues and scattering eigenvectors can be expressed as $\lambda^{(s,p)}_{\pm}=\frac{\sqrt{1-M^{(s,p)}_{11}M^{(s,p)}_{22}}}{M^{(s,p)}_{22}}\cos\phi^{(s,p)}_R\pm \sqrt{\frac{M^{(s,p)}_{11}M^{(s,p)}_{22}+\cos^2\phi_R^{(s,p)} [1-M^{(s,p)}_{11}M^{(s,p)}_{22}]}{M^{(s,p)2}_{22}}}$ and $\vert S_{\pm}>=[-\frac{1}{M^{(s,p)}_{22}},
-i\frac{\sqrt{1-M^{(s,p)}_{11}M^{(s,p)}_{22}}}{M^{(s,p)}_{22}}\sin\phi_R^{(s,p)}\mp \sqrt{\frac{1}{M_{22}^{(s,p)2}}-\frac{1-M^{(s,p)}_{11}M^{(s,p)}_{22}}{M_{22}^{(s,p)2}}\sin^{2}\phi^{(s,p)}_R} ]^T$, respectively.
By analyzing $\lambda^{(s,p)}_{\pm}$ and $\vert S^{(s,p)}_{\pm}>$, the scattering phases can be inferred.
In symmetry phase, the scattering eigenvalues $\lambda^{(s,p)}_{\pm}$ are two distinct reals, resulting in amplification or dissipation in outgoing waves.
Moreover, input intensities corresponding to   each scattering eigenvectors
are same.
On contrary, in broken symmetry phase, two scattering eigenvalues are same in magnitudes, i.e., $\vert \lambda^{(s,p)}_{+}\vert =\vert \lambda^{(s,p)}_{-}\vert $, while the input intensity ratio with each scattering eigenvectors can establish a pair with reciprocal moduli.
An EP appears in $(r^{(s,p)}_L-r^{(s,p)}_R)^2+4t^{(s,p)2}=0$.
Again, with parametrization, we can depict these scattering phases and EP by various $\phi^{(s,p)}_R$ in the parametric space of Fig. 1 (c).
Here the purple lines represent an EP, while the blue and green color regions are denoted by symmetry broken phase and symmetry phase, respectively.
We should mention that the EP of APT systems allows various $R$ and $T$.
By comparison, an EP in PT-symmetric systems exhibits the characteristic of $R=0$ and $T=1$ \cite{PT2,PT6,PT7}.

 \begin{figure}[ht]
 \centering
\includegraphics[width=0.4\textwidth]{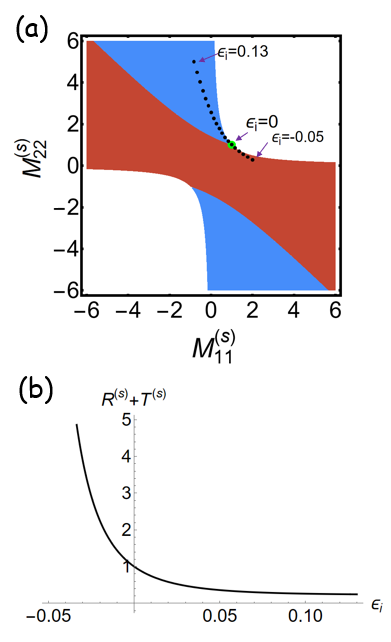}
\caption{A bilayer APT system made of $l=1.775\lambda_0$,$\mu=1.56$,  $\epsilon=2.3+i\epsilon_i$, and $\theta_{in}=1.19$ rad is considered. As we tune $\epsilon_{i}$, we analyze their parametrization in (a), while the corresponding $R^{(s)}+T^{(s)}$ is shown in (b). }
\end{figure}  

\subsection{Parametric space for  APT-symmetric bilayer systems with lossless, loss, and gain media}
The imaginary part of $n(x)$ in arbitrary bilayer  APT systems can have three situations, namely, all positive (loss), all negative (gain), or zero (lossless).
Moreover, from power conservation, we argue that $T+R<1$, $T+R>1$, and $T+R=1$ can be established in  lossy, gain, and lossless, respectively.

To support the result, we discuss a APT bilayer system made of  $l=1.775\lambda_0$, $\mu=1.56$,$\epsilon=2.3+i\epsilon_i$, and $\theta_{in}=1.19 rad$.
Here $\lambda_0$ is operating wavelength.
Under TE polarization incidence,   the parametrization is shown in Fig 2. (a), while we tune $\epsilon_i\in[-0.05,0.13]$.
We can see that as $\epsilon_i\in[-0.05,0)$, $\epsilon_i\in(0,0.13]$, and $\epsilon_i=0$, wave  in associated systems experiences amplification with $T^{(s)}+R^{(s)}>1$, dissipation with $T^{(s)}+R^{(s)}<1$, and lossless propagation with $T^{(s)}+R^{(s)}=1$, respectively, as shown in Fig. 2 (b).

  \section{Omnidirectional wave impedance matching}
We utilize wave impedance analysis to characterize any  bilayer APT systems.
We first derive a propagation matrix $M^{(WI)}$ for arbitrary bilayer systems having APT property.
Under TE polarization incidence, it reads
\begin{equation}\label{bilayer}
\begin{split}
 \begin{bmatrix}
 E^{(s)}_{y}(l)\\
 H^{(s)}_{x}(l)
 \end{bmatrix}
 &=M^{(WI,s)}\begin{bmatrix}
 E^{(s)}_y(-l)\\
 H^{(s)}_x(-l)
 \end{bmatrix}
\end{split}
\end{equation}
where 
\begin{equation}
\begin{split}
\begin{cases}
M^{(WI,s)}_{11}&=\cos[\nu l]\cos[\nu^{*}l]+\frac{\nu}{\nu^{*}}\sin[\nu l]\sin[\nu^{*}l]\\
M^{(WI,s)}_{12}&=-i\{\frac{\mu \omega}{\nu}\cos[\nu^{*} l]\sin[\nu l]-\frac{\mu \omega}{\nu^{*}}\cos[\nu l]\sin[\nu^{*} l]\}\\
M^{(WI,s)}_{21}&=-i\{\frac{\nu}{\mu \omega}\cos[\nu^{*} l]\sin[\nu l]-\frac{\nu^{*}}{\mu \omega}\cos[\nu l]\sin[\nu^{*} l]\}\\
M^{(WI,s)}_{22}&=\cos[\nu l]\cos[\nu^{*}l]+\frac{\nu^{*}}{\nu}\sin[\nu l]\sin[\nu^{*}l]\\
\end{cases}
\end{split}
\end{equation} 
here $E^{(s)}_{y}$ and $H^{(s)}_x$ are denoted as y-component of total electric field and  x-component of total magnetic field, respectively, while $\nu$ is  z-component of wave number at the left slab, i.e., $k_{1z}=k_0\sqrt{\mu\epsilon-\sin^2\theta_{in}}\equiv \nu$, and $k_0$ is wave number of free space.
The detailed derivation is placed in Appendix B.
Utilizing lossless $\epsilon$ to the propagation matrix $M^{(WI,s)}$, we obtain
\begin{equation}
\begin{split}
\begin{bmatrix}
 E^{(s)}_{y}(l)\\
 H^{(s)}_{x}(l)
 \end{bmatrix}
=\begin{bmatrix}
 1 & 0\\
 0 & 1
 \end{bmatrix}\begin{bmatrix}
 E^{(s)}_y(-l)\\
 H^{(s)}_x(-l)
 \end{bmatrix}.
 \end{split}
\end{equation}
Interestingly, $M^{(WI,s)}$ becomes a unit matrix.
Accordingly, it represents $t^{(s)}=1$ and $r^{(s)}_L=r^{(s)}_R=0$.
By the parametrization, it corresponds to $M^{(s)}_{11}=1$ and $M^{(s)}_{22}=1$, as marked by a red dot in Figs. 1 (c)-(d).

On the other hand, based on  Eq.(\ref{bilayer}), we derive a relation for wave impedances at $z=-l$ and $z=l$,
\begin{equation}\label{boundary}
Z^{(s)}(l)=\frac{M^{(WI,s)}_{12}-Z^{(s)}(-l)M^{(WI,s)}_{11}}{M^{(WI,s)}_{21}Z^{(s)}(-l)-M^{(WI,s)}_{22}}.
\end{equation}
Suppose that there exists  reflectionless, the corresponding wave impedances at $z=-l$ and $z=l$ should meet $Z(-l)=\frac{\eta_0}{\cos\theta_{in}}$ and $Z(l)=\frac{\eta_0}{\cos\theta_{in}}$ , respectively. 
Here $\eta_0=\sqrt{\frac{\mu_0}{\epsilon_0}}$ is intrinsic impedance of free space.
Taking these wave impedance boundary conditions into Eq.(\ref{boundary}), we have
\begin{widetext}
\begin{equation}\label{reflectionless}
\begin{split}
2iIm[\frac{\nu^{*}}{\nu}]\vert \sin(\nu l)\vert^2
&=2 Re[i\cos[\nu^{*}l]\sin[\nu l]\frac{ \mu k_0 \cos[\theta_{in}] }{\nu }]-2Re[\frac{i \nu }{\mu k_0\cos[\theta_{in}]}\cos[\nu^{*}l]\sin[\nu l]].
\end{split}
\end{equation}
\end{widetext}
Remarkably,  the left-hand expression is purely imaginary, whereas the right-hand one is purely real.
This seemingly contradiction can be resolved only when $\epsilon$ and $\mu$ are real.
Accordingly, both left-hand and right-hand expressions become zero, resulting in WIM.
Interestingly, the formation of WIM is irrespective of $l$,  any values of refractive indexes, and  incident angles.
Furthermore, under TM polarization incidence, we also find that a WIM covering all incident angles can also be established, as the APT system is lossless.
The formulation is placed in Appendix B.
We further refer to this remarkable result as polarization-independent omnidirectional wave impedance matching (polarization-independent OWIM), i.e., wave impedance matching that holds for all incident angles and polarizations.
On the other hand, Eq.~(\ref{reflectionless}) implies that when gain or loss is introduced into an APT system, the formation of polarization-independent OWIM is disrupted, resulting in wave reflection.

To illustrate the polarization-independent OWIM phenomenon, an electric dipole, whose its  dipole moment is along $y$, is placed in vicinity of a lossless bilayer APT system, as shown in Fig. 3 (a).
For this system, we let $\mu=0.01$,$\epsilon=0.01$, and $ l=5\lambda$.
For comparison, we also consider the same electric dipole placed in a free space in Fig. 3 (b). 
It confirms the formation of polarization-independent OWIM.
However, as for PIM or NIM alone, a total internal reflection occurs at an incident angle beyond the associated critical angle, as in Fig. 3 (c).
We remark that the polarization-independent OWIM appeared in our lossless APT-symmetric systems is a generalization of the classic example of a flat slab with $\epsilon=-1$ and $\mu=-1$ \cite{APT1}.

 \begin{figure}[ht]
 \centering
\includegraphics[width=0.5\textwidth]{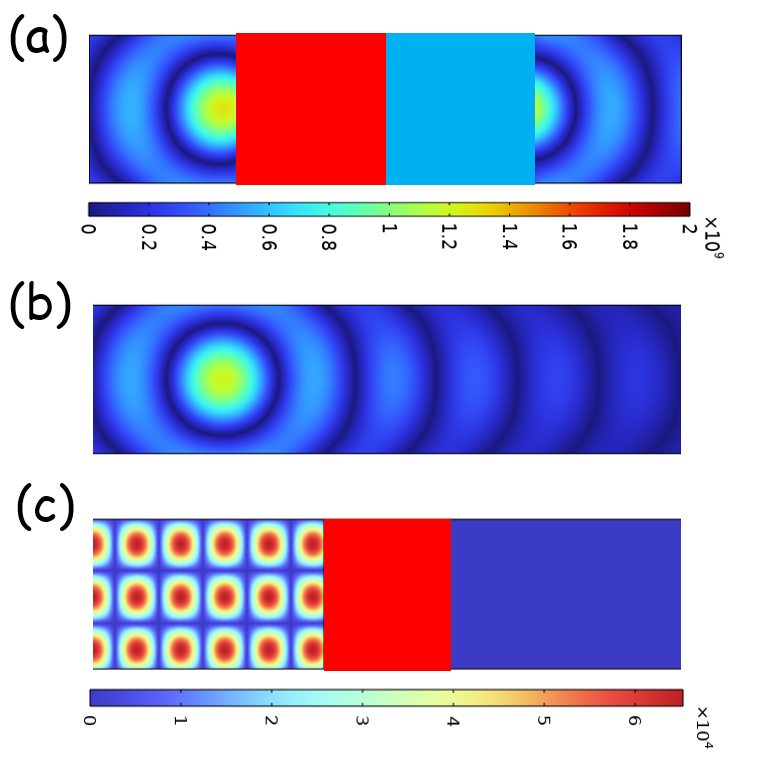}
\caption{Electric field distributions of an electric  dipole placed in close to a lossless APT bilayer system and in free space are shown in (a) and (b), respectively. The red and blue boxes represent the bilayer system made of PIM and NIM. (c) As for a PIM's slab alone, there occurs a total internal reflection  as their incident angle is beyond the critical one. In this APT system, we use $\mu=0.01$, $\epsilon=0.01$,  $l=5\lambda_0$, and $\theta_{in}=\pi/4 $ rad. Here, the color bars represent the magnitude of the electric field in units of $\mathrm{V/m}$.}
\end{figure}

  \begin{figure*}[ht]
 \centering
\includegraphics[width=1\textwidth]{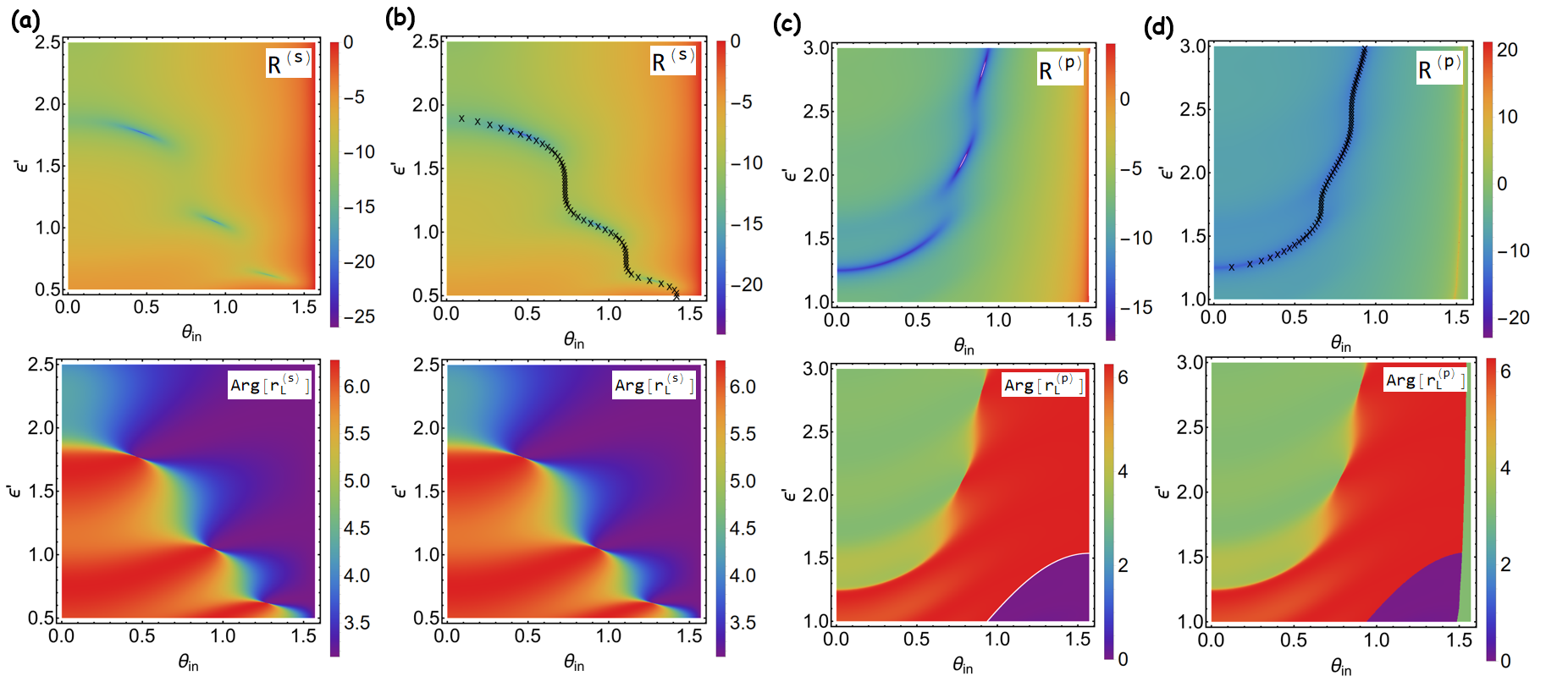}
\caption{Contour plots of reflectances for left incidence and their associated reflection phase with respect to incident angles $\theta_{in}$ and real parts of $\epsilon$ are shown in the top and bottom panels, respectively. 
In (a) and (b), they are obtained from Eq.(7) and COMSOL simulation, respectively, while
in (c) and (d), they  are obtained from Eq.(9) and COMSOL simulation, respectively.}
\end{figure*} 

\section{Reflection dips}
In non-Hermitian domain, we can express relative permittivity by $\epsilon=\epsilon^{'}+i\epsilon^{''}$ and z-component of wave number in the left slab by $\nu=\nu^{'}+i\nu^{''}$, where $\epsilon^{'}$, $\epsilon^{''}$, $\nu^{'}$, and $\nu^{'}$ are real.
$\nu^{''}$ is responsible for lossy when $\nu^{''}>0$ or for gain when $\nu^{''}<0$.
Given that lossless APT systems exhibit wave impedance matching for all incident angles and polarizations, we consider the influence of a small amount of loss or gain, i.e., $\vert \epsilon^{''}\vert <<1$. 
We thus derive $\nu^{'}=\sqrt{k_0^2\mu\epsilon^{'}-k_0^2\sin^2\theta_{in}}$ and $\nu^{''}=\frac{k_0^2\mu\epsilon^{''}}{2\nu^{'}}$.
Accordingly, we obtain the reflection coefficient for left incidence under TE polarization,
\begin{widetext}
\begin{equation}
r_L^{(s)}\approx -2i\frac{\nu^{''}}{\nu^{'}}\sin^2[\nu^{'}l]+\cos\theta_{in}\mu_r k_0\frac{\nu^{''}}{2\nu^{'2}}(2\nu^{'}l-\sin[2\nu^{'}l])-\frac{\nu^{''}}{2\cos\theta_{in}\mu_r k_0}(2\nu^{'}l+\sin[2\nu^{'}l]).
\end{equation}
\end{widetext}
The derivation  is given in Appendix C.
Additionally, we note that with APT condition, it satisfies $r_L^{(s,p)}=r_R^{(s,p)*}$.

Based on Eq.(7), we provide  contour plots for the logarithm of reflectance $Ln(R^{(s)})$ and the reflection phase $Arg[r^{(s)}_L]$ by scanning incident angles $\theta_{in}$ and real parts of $\epsilon$, that is, $\epsilon^{'}$, shown in the top and bottom panels of Fig. 4 (a), respectively.
Here we consider a lossy APT bilayer system made of $\mu=2.1,l=0.8\lambda_0,Im[\epsilon]=0.01$.
By comparison,  we also provide associated calculation but with COMSOL simulation in Fig. 4 (b), supporting the accuracy of our analytical expression in Eq. (7).
Additionally, when $\epsilon^{''}=0$, we have $r_L^{(s)}=0$ in Eq.(7), consistent with our previous statement. 
Moreover, we can observe that reflection dips, i.e., minimum reflectance, are accompanied by a jump of reflection phase.
To understand the underlying mechanism, we analyze Eq.(7),
which is constituted by the real and the imaginary parts.
There are two terms in the real part. 
We can argue that the reflection phase jump is caused by the sign change of the real part term. 
Accordingly, we obtain the condition of a minimum $R^{(s)}$, 
\begin{equation}
(k_0\mu_r\cos\theta_{in})^2=\nu^{'2}\frac{\nu^{'}l+\sin[\nu^{'}l]\cos[\nu^{'}l]}{\nu^{'}l-\sin[\nu^{'}l]\cos[\nu^{'}l]}.
\end{equation} 
With this relation, we numerically solve $\epsilon^{'}$ in terms of $\theta_{in}$, as marked by black arrows in the top panel of Fig. 4 (b), confirming the location of reflection dips.

\begin{figure*}[ht]
 \centering
\includegraphics[width=1\textwidth]{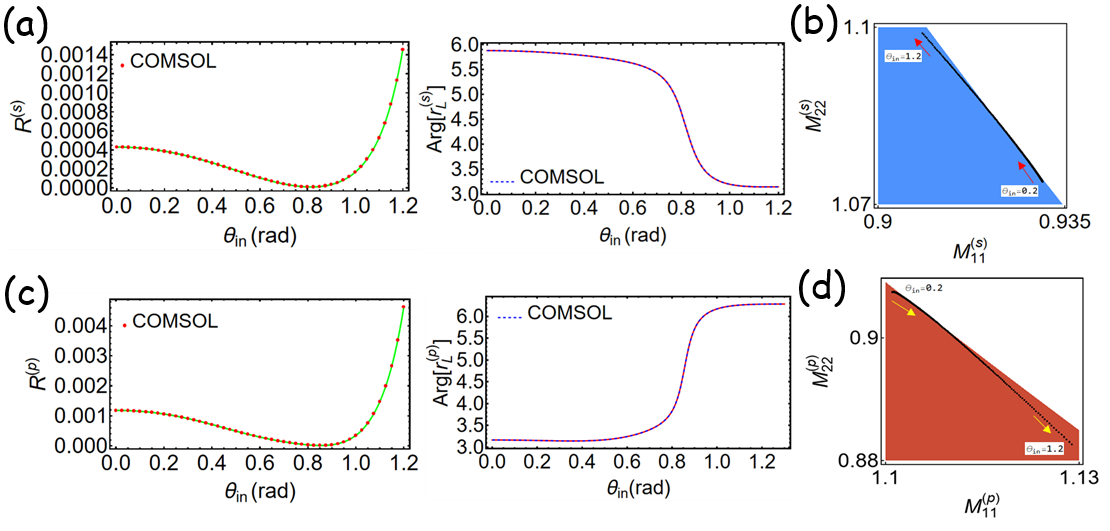}
\caption{Reflectances and their reflection phases with respect to $\theta_{in}$ in (a) for TE polarization incidence and (c) for TM polarization incidence. Here the green lines are obtained from Eqs.(7) and (9), while the red dots are obtained from COMSOL. A reflection dip is accompanied by a reflection phase jump. With the parametric space, we analyze these systems in (b) and (d), corresponding to (a) and (b), respectively. }
\end{figure*} 

Next, we consider TM polarization incidence.
The corresponding reflection coefficient for left incidence reads
\begin{widetext}
\begin{equation}
\begin{split}
r_L^{(p)}
&\approx 2i\sin^2[\nu^{'}l](\frac{\epsilon_r^{''}}{\epsilon_r^{'}}-\frac{\nu^{''}}{\nu^{'}})-\frac{k_0\cos\theta_{in}}{2\nu^{'2}}[-2l\epsilon_r^{'}\nu^{'}\nu^{''}-\sin[2\nu^{'}l](\nu^{'}\epsilon_r^{''}-\epsilon_r^{'}\nu^{''})]-\frac{1}{2k_0\cos\theta_{in}\epsilon_r^{'}}[2l\nu^{'}\nu^{''}-\sin[2\nu^{'}l](\nu^{'}\frac{\epsilon_r^{''}}{\epsilon_r^{'}}-\nu^{''})].
\end{split}
\end{equation}
\end{widetext}
The derivation is given in Appendix C.
Again, as $\epsilon^{''}=0$, we have $r_L^{(p)}=0$ as expected.
Now,  we employ a gain APT bilayered system made of $\mu=1.3$, $l=2\lambda_0$, and $Im[\epsilon]=-0.01$.
Based on Eq.(9), we calculate the corresponding contourplots for the logarithm of reflectance $Ln(R^{(p)})$ and associated reflection phase $Arg[r_L^{(p)}]$ in the top and bottom panels of Fig. 4 (c), respectively.

Meanwhile, we provide associated numericals by COMSOL simulation in Fig. 4 (d), which they match very well. 
In Figs. 4 (c)-(d), we also observe that the occurrence of reflection dips is accompanied by reflection phase jumps.
Here, it is from $\pi$ to $2\pi$.
We state that the cause of a reflection phase jump is due to the sign change of $Re[r_L^{(p)}]$.
Thus, we derive the condition of minimum $R^{(p)}$ 
 \begin{equation}
 (k_0\cos\theta_{in})^2=\frac{\nu^{'2}[2l\nu^{'}\nu^{''}-\sin[2\nu^{'}l](\nu^{'}\frac{\epsilon_r^{''}}{\epsilon_r^{'}}-\nu^{''})]}{\epsilon_r^{'2}[2l\nu^{'}\nu^{''}+\sin[2\nu^{'}l](\nu^{'}\frac{\epsilon_r^{''}}{\epsilon_r^{'}}-\nu^{''})]}.
 \end{equation}
 Accordingly, we numerically solve  $\epsilon^{'}$ in terms of $\theta_{in}$ as illustrated by black arrows in Fig. 4 (d), consistent with the locations of reflection dips.

\subsection{Examples}
To verify our finding of the reflection dip in non-Hermitian bilayer APT systems, we provide two systems. 
Following  the result of Figs. 4 (a)-(b), we  present a lossy system made of $l=0.8\lambda_0$, $\mu=2.1$, and $\epsilon=1.14+0.01i$, while we calculate 
the reflectance for left incidence and associated reflection phase with respect to incident angles $\theta_{in}$ under TE polarization incidence in Fig. 5 (a).
We can observe that there has a reflection dip at $\theta_{in}=0.815$ rad,  with  $R^{(s)}=1.08\times 10^{-5}$.
Notably, the reflection phase exhibits a jump from $2\pi$ rad to $\pi$ rad around $\theta_{\mathrm{in}} = 0.8$ rad.
However, the reflection phase by the right incidence would become $\pi$ rad to $2\pi$ rad, due to APT.
Next, we discuss the parametric space for this system,  as shown in Fig. 5 (b).
The system lies in the lossy domain as expected.
Meanwhile, the parametric trajectories are close to the boundary of reflectionless, corresponding to evidence of a low reflectance.

Next, we consider a gain APT system under TM polarization incidence.
Here we exploit $l=2\lambda_0$, $\mu=1.3$, and $\epsilon=2.45-0.01i$.
In Fig. 5 (c), we provide the corresponding reflectance $R^{(p)}$ and reflection phase $Arg[r_L^{(p)}]$. 
We can observe that the reflection dip occurs at  $\theta_{in}=0.85$ rad, with $R_L^{(p)}=1.23\times 10^{-5}$.
Again, we employ the parametric space to analyze this system.
As shown in Fig. 5 (d), the parametric trajectories lie in the gain domain as expected, while they are close to the boundary of reflectionless.

\begin{figure*}[ht]
 \centering
\includegraphics[width=1\textwidth]{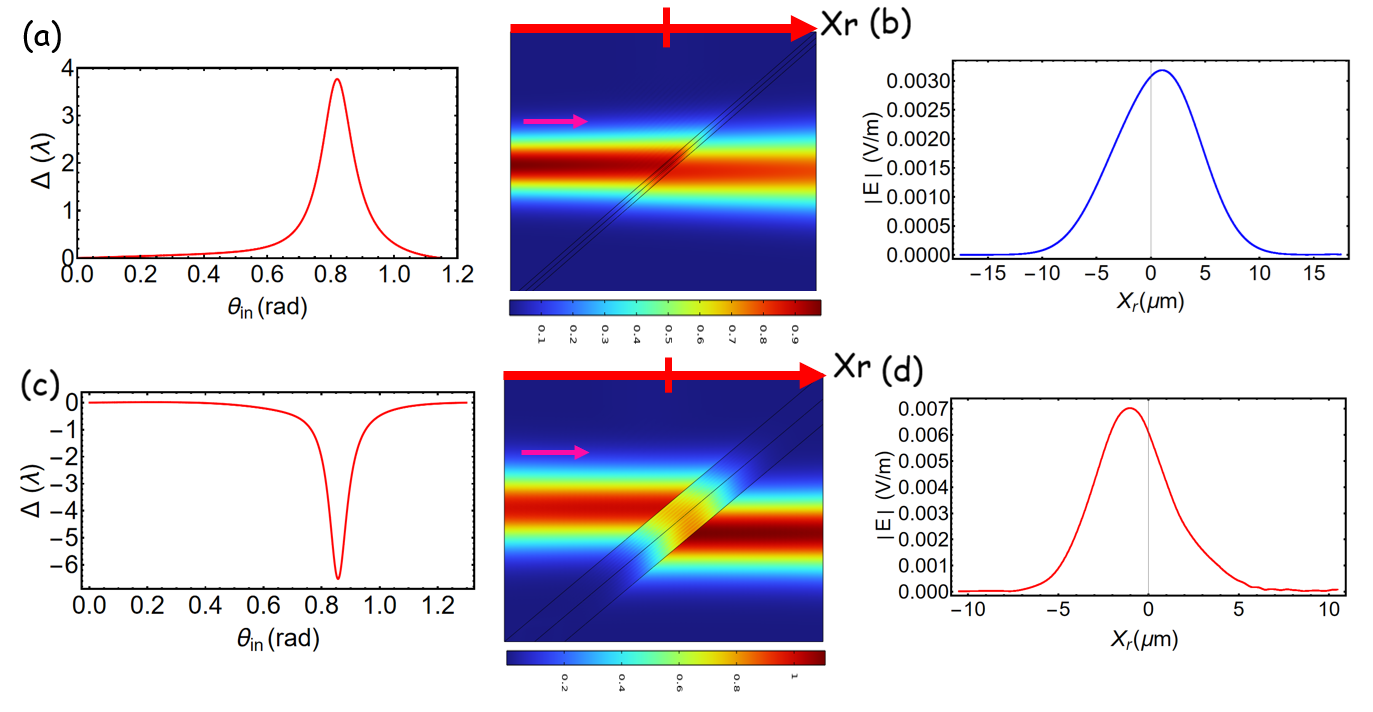}
\caption{The GH shifts of the systems in Figs. 5(a) and 5(c) are calculated in (a) and (c), respectively. A TE- and a TM-polarized Gaussian beams are incident upon the corresponding APT systems in the right panels of (a) and (c), respectively. Here the pink arrows indicate the direction of the incident beam, while the color bars represent the magnitude of the electric field in units of $\mathrm{V/m}$. The  electric field intensities of the reflection beams along $X_r$ axis are evaluated in (b) and (d), corresponding to the results demonstrated in the right panels of (a) and (c). }
\end{figure*}

\section{Goos-H\"anchen Effect}
As the incident field is a beam packet, the reflection beam would encounter a longitudinal shift as well as a transverse shift, which cannot be described by geometric optics \cite{GH1,GH2,GH3,IF1,IF2}.
The former effect is known as the Goos–H\"anchen (GH) shift, first reported in total internal reflection, which occurs within the incidence plane \cite{GH1,GH2}. 
This effect is attributed to the fact that each plane wave component, whose wave vectors slightly deviate from the beam coordinate, undergoes different reflection phases.
The latter one is referred to as the spin Hall effect of light, also known as the Imbert-Fedorov effect, predicted by F.I. Fedorov in 1955 \cite{IF1}. 
In this effect, the beam shift occurs perpendicular to the incidence plane and can be attributed to the conservation of the total angular momentum of light accounting for orbital momentum and  spin momentum \cite{spin1,spin2}.

Although the GH shift has been extensively investigated in weakly dissipative media \cite{GH4,GH5}, metal \cite{GH51}, negative index media \cite{GH6}, epsilon-near-zero metamaterials \cite{GH7,GH71,GH72},  photonic crystals \cite{GH8,GH9,GH10}, PT symmetry systems \cite{PTcrytal1,GH12}, Weyl semimetals \cite{GH13,GH14},  topological materials \cite{GH141} to name a few, we find that the GH effect related to APT systems remains unexplored.

 \begin{figure*}[ht]
 \centering
\includegraphics[width=1\textwidth]{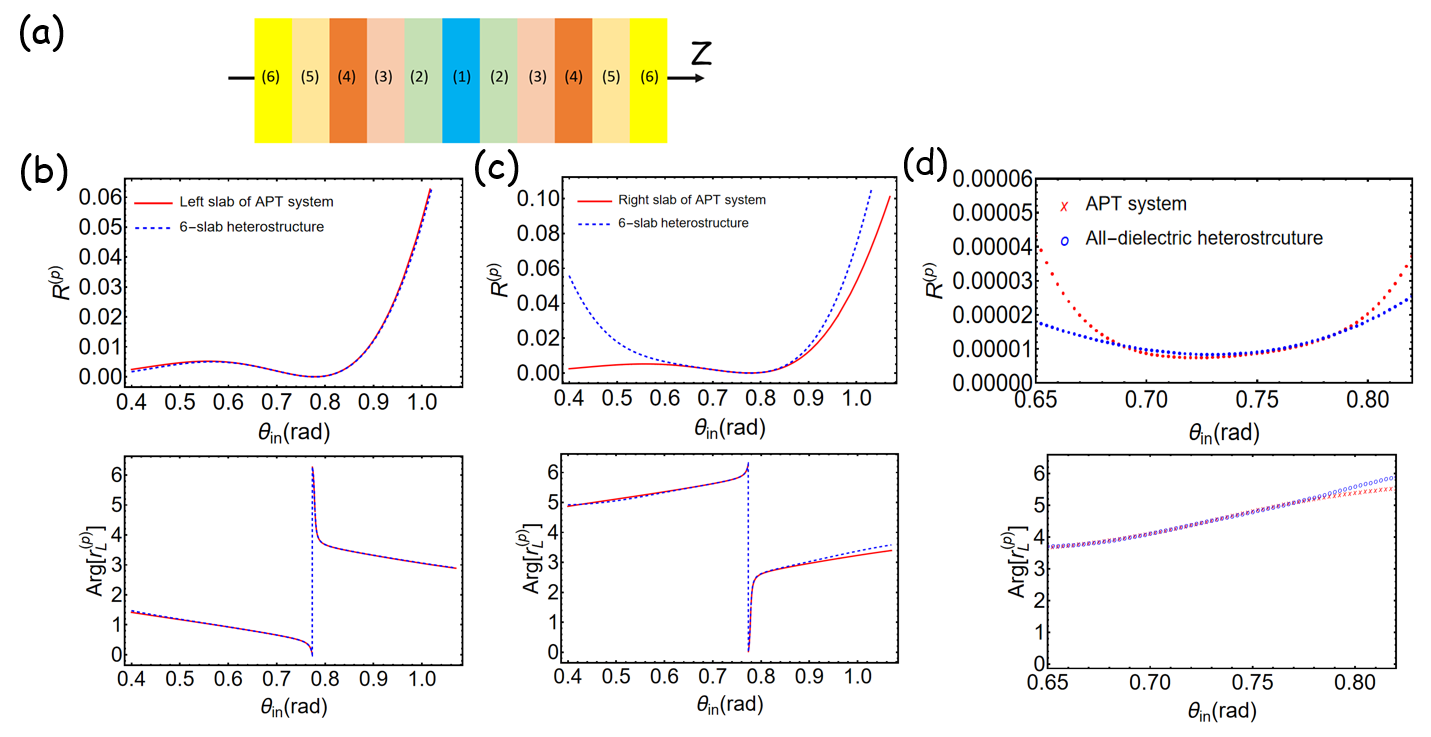}
\caption{(a) Schematic of a  mimicking heterostructure with six slabs, where the index denotes the slab order. 
Reflectance and reflection phase for the two mimicking heterostructures and the corresponding component slabs of the bilayer APT system are shown in (b) and (c). 
(b) mimics the left slab, whereas (c) mimics the right slab.
 The blue dashed curves represent the  results calculated from each all-dielectric heterostructure, while the red curves correspond to those component slabs of the APT system. By combining the two heterostructures into a composite system, the corresponding reflectance and reflection phase are calculated and shown by the blue circles in (d), whereas the red crosses denote those of the mimicked APT system.}
\end{figure*}

Following Artmann's formulation for a total internal reflection \cite{GH15}, the description of the GH shift, i.e., a longitudinal shift, is given by
 \begin{equation}
 \triangle^{(s,p)}=-\frac{1}{k_0\cos\theta_{in}}\frac{d\phi^{(s,p)}_r}{d\theta}\vert_{\theta_{in}},
 \end{equation}
  indicating that a variation of the reflection phase with respect to $\theta_{in}$ results in the GH effect.
We should note that the GH shift is polarization dependence.
Though this expression is originally derived within the framework of a total internal reflection, that is, the stationary-phase method, it remains valid for partial reflections, as well explored in Brewster angle effect and PT-symmetry crystals \cite{PTcrytal1,GH16,GH5}.
Now, by the results in Figs. 5 (a) and 5 (c), we evaluate the corresponding GH shift as shown in Figs. 6 (a) and 6(c), respectively.
As expected, a significant GH shift occurs at the jump of reflection phase, although both cases have low reflectance.
To further verify the GH phenomenon, we perform COMSOL simulations, as shown in the right panels of Figs. 6 (a) and 6 (c), where the incident fields are TE-polarized and TM-polarized Gaussian beams, respectively.
The central wavelengths of the Gaussian beams in Figs. 6 (b) and 6 (d) are both $\lambda_0 = 700$ nm.
We observe that both transmission beams are dominant, while reflection beams have almost vanished.
Moreover, the transmission beams are attenuated and amplified in Figs. 6 (a) and (c), owing to loss and gain in APT systems.
Especially, the transmitted beam in NIM exhibits negative refraction, as clearly shown in the right panel of Fig. 6 (c).
Next, we evaluate the electric field intensity of the reflection beams along  $X_r$ axis, as shown in the right panels of Figs. 6 (b) and 6 (d), where the positive and negative shifts are observed, respectively.
As a consequence of APT condition, the GH shift on the opposite side is equal in magnitude but opposite in sign.
Last but not least, the enhanced GH effect is facilitated by a phase jump occurred at a reflection dip, analogous to the Brewster angle effect, as well stated in Ref. \cite{GH16,GH5}.

\section{Emulation of Electromagnetic Response in Bilayer APT Systems Using an All-Dielectric Heterostructure}
Interestingly,  the NIM, regarded as a hallmark of metamaterials, has been found in Dirac semimetals  \cite{effective1}.
However, realizing APT systems generally requires stringent constraints on the electromagnetic materials, which may be difficult to find in natural materials.
Instead, we seek an effective medium theory together with a simple implementation architecture.
Recently, it has been demonstrated that one dimensional photonic crystals with unit cells composed of spatially symmetric dielectric constituents can exhibit the electromagnetic response of negative permittivity \cite{effective2}.
More recently, with proper structural designs, all-dielectric  heterostructures can emulate the same electromagnetic fields of APT systems \cite{APT3}. 
The approach is to require an all-dielectric symmetric heterostructure having the same transmission and reflection as the target slab.

Now, we adopt an all-dielectric heterostructure to mimic the electromagnetic angular dispersion of APT  systems within a finite frequency window.
As demonstration, we consider a gain bilayer APT system. 
For the left slab of the gain APT system, we employ $\lambda_0=600$ nm, $l=1.22\lambda$, $\epsilon_L=2.1-0.01i$, and $\mu_L=1.3$.
We calculate the reflectance and associated reflection phase  with $\theta_{in}$ under a TM-polarized incidence  as shown in the red lines Fig. 7 (b). 
We can observe that there is a reflection dip at $\theta_{in}=0.78$, while it is also accompanied by a reflection phase jump from $0$ rad to $2\pi$ rad.
Now, we employ an all-dielectric heterostructure with six slabs, whose the structure is symmetry in space as shown in Fig. 7 (a).
Then we choose each thickness parameters with $l_1=110[nm]$, $l_{2}=120$ nm, $l_3=100$ nm, $l_4=30$ nm, $l_5=30$ nm, and  $l_6=31$ nm.  
Next, we require this all-dielectric symmetric heterostructure to have the same transmission and reflection as the left slab at $\theta_{in}=[0.7 \mathrm{rad},0.73 \mathrm{rad}, 0.77 \mathrm{rad}]$. 
As a result, we numerically find that to construct the mimicking all-dielectric heterostructure, it needs
$\epsilon_{1}=4.09-0.89 i$,
$\epsilon_{2}=3.65 + 0.99i$,
$\epsilon_{3}=2.02 - 0.39i$,
$\epsilon_{4}=2.24+ 0.76i$,
$\epsilon_{5}=1.56 -1.64i$, and
$\epsilon_{6}=1.12 +0.28i$.
We calculate the corresponding reflectance and
associated reflection phase within $0.4\mathrm{rad} <\theta_{in}<1.07 \mathrm{rad}$ as shown in the blue dashed lines of Fig. 7 (b), whose their angular dispersion of reflection matches that of the left slab.

Next, we consider the right slab of the target APT system.
The corresponding reflectance and reflection phase are given by the red lines in Fig. 7 (c). 
Again, we employ another all-dielectric heterostructure made of six slabs whose spatial placement is in a symmetric manner.
We use the same thickness parameters as that in Fig. 7 (b), but with different material parameters.
Following the same  approach, we numerically find that the electromagnetic material parameters given by $\epsilon_{1}=6.06 +14.19i$,
$\epsilon_{2}=2.43 -0.77i$,
$\epsilon_{3}=-0.86 -0.025i$,
$\epsilon_{4}=13.31 + 1.43i$,
$\epsilon_{5}=0.03 -1.26i$, and
$\epsilon_{6}=12.97 +8.58i]$ can meet the conditions.
The calculated reflectance and reflection phase of the all-dielectric heterostructure are given by the blue dashed curves in Fig. 7 (c).
Even small deviations for two reflectances can occur at $\theta_{in}<0.6 \mathrm{rad}$ and $\theta_{in}>0.9 \mathrm{rad}$, we can state that two systems have nearly identical angular dispersion.
Next, by combining the two all-dielectric heterostructures into a composite system, we evaluate the corresponding reflectance and reflection phase shown in the blue circles of Fig. 7 (d). 
For comparison, the reflectance and reflection phase of the gain  APT system are presented by the red crosses in Fig. 7 (d). 
We can see that, as $\theta_{in}\in [0.68 \mathrm{rad}, 0.79 \mathrm{rad}]$, two systems perform nearly identical dispersion relation of reflection.

So far we have mimicked electromagnetic responses from the target APT system by all-dielectric heterostructures made of a total of eleven slabs in a range of incident angles.
We expect that, with a complexity of slabs involved, the mimicked electromagnetic response can be improved over a broader range of incident angles.
Last but not least, we emphasize that polarization-independent OWIM realized in balanced and lossless PIM–NIM metamaterials offers greater flexibility and simplicity than more complex artificial structures, such as spatiotemporally modulated media, anisotropic metamaterials, Chebyshev transformer metasurfaces, and photonic crystals \cite{OMN1,OMN2,OMN3,OMN4,OMN5}.
Furthermore, the required material condition for polarization-independent OWIM is scalable across different frequency ranges and irrespective of system size.

\section{Conclusions}
In summary, we consider electromagnetic wave reflection from arbitrary bilayer APT systems.
With the parametric space as well as wave impedance analysis, any lossless APT systems enable WIM for all incident angles and incident polarization, irrespective of  system's size and refractive index.
In the presence of a small gain or loss, the WIM result can not be satisfied; instead, a reflection dip emerges.
We analytically derive a closed form for  reflection.
The reflection dip is accompanied by a jump of reflection phase.
By leveraging the phase jump, we discuss the GH effect enhanced, which remains unexplored in APT systems.
To mimic the electromagnetic field generated from the APT systems, we employ all-dielectric symmetric heterostructure to have the same electromagnetic response of each slab components over a specific angular spread.
Our work may inspire a variety of applications, including optical sensing, subwavelength imaging, precise metrology, spin-based devices, radar stealth, and invisibility cloaking.

 \section{Appendix A}
Consider an APT system consisting of an even number $N$ of slabs, where the refractive index, relative permeability, and thickness of the $i$th slab are denoted by $n_i$, $\mu_{r,i}$, and $l_i$, respectively, with $i = 1, \ldots, N$.

Under TE (i.e., $\mathbf{s}$) polarization, we derive the transfer matrix for the interface between free space (on the right) and the first slab (on the left):
 \begin{equation}
 \begin{split}
 M^{(s)}_{free\rightarrow n_1}=\frac{1}{2}\begin{bmatrix}
 1+\frac{\mu_1 k_{0z}}{\mu_0 k_{1z}} & 1-\frac{\mu_1 k_{0z}}{\mu_0 k_{1z}}\\
  1-\frac{\mu_1 k_{0z}}{\mu_0 k_{1z}} & 1+\frac{\mu_1 k_{0z}}{\mu_0 k_{1z}}
 \end{bmatrix}
 \end{split}
 \end{equation}
 where $k_0$ is wave number in free space, $k_{0z}=\sqrt{k_0^2-k_0^2\sin^2\theta_{in}}$ is z-component of wave number in free space and $k_{1z}=\sqrt{k_1^2-k_0^2\sin^2\theta_{in}}$ is z-component of wave number  in the first slab with $[\epsilon_1,\mu_1,l_1]$.
 
At an interface between free space on the left sideand the $N$-th slab on the right side, the corresponding transfer matrix can be expressed as
  \begin{equation}
 \begin{split}
 M^{(s)}_{ n_N\rightarrow free}=\frac{1}{2}\begin{bmatrix}
 1+\frac{\mu_0 k_{Nz}}{\mu_N k_{0z}} & 1-\frac{\mu_0 k_{Nz}}{\mu_N k_{0z}}\\
  1-\frac{\mu_0 k_{Nz}}{\mu_N k_{0z}} & 1+\frac{\mu_0 k_{Nz}}{\mu_N k_{0z}}
 \end{bmatrix}
 \end{split}
 \end{equation}
here $k_{Nz}=\sqrt{k_N^2-k_0^2\sin^2\theta_{in}}$ is z-component of wave number in the N-th slab with $[\epsilon_N,\mu_N,l_N]$.\\

At the interface of each slabs, the transfer matrix can be expressed as 
  \begin{equation}
 \begin{split}
 M^{(s)}_{ n_i\rightarrow n_{i+1}}=\frac{1}{2}\begin{bmatrix}
  1+\frac{\mu_jk_{iz}}{\mu_i k_{jz}} & 1-\frac{\mu_jk_{iz}}{\mu_i k_{jz}}\\
  1-\frac{\mu_jk_{iz}}{\mu_i k_{jz}} & 1+\frac{\mu_jk_{iz}}{\mu_i k_{jz}}
 \end{bmatrix}
 \end{split}
 \end{equation}
 where $i=1, \ldots, N-1$ and $k_{iz}=k_0\sqrt{n_i^2-\sin^2\theta_{in}}$ is z-component of wave number in the i-th slab.
The transfer matrix for wave propagation in each slab is
 \begin{equation}
 \begin{split}
 M_{n_i,l_i}=\begin{bmatrix}
 e^{ik_{iz}l_i} & 0\\
 0&   e^{-ik_{iz}l_i}
 \end{bmatrix}.
 \end{split}
 \end{equation}

  Finally, the transfer matrix for a system consisting of $N$ slabs is given by
 \begin{equation}
 \begin{split}
 M_{tot}^{(s)}=M^{(s)}_{ n_N\rightarrow free} M_{n_N,l_N}[\Pi_{i=1}^{i=N-1} M^{(s)}_{ n_i\rightarrow n_{i+1}}M_{n_i,l_i}]M^{(s)}_{free\rightarrow n_1}.
 \end{split}
 \end{equation}
 
Now,  under a spatial inversion with respect to the origin of coordinate, i.e., $-z\rightarrow z$, the APT condition requires $n\rightarrow -n^{*}$ and $\mu\rightarrow -\mu$. 
So we have
\begin{equation}
\begin{split}
&\sqrt{n^2-\sin^2\theta_{in}}=\sqrt{(n_r^2-n_i^2-\sin^2\theta_{in})+2in_rn_i}\\&\rightarrow \sqrt{(n_r^2-n_i^2-\sin^2\theta_{in})-2in_rn_i},
\end{split}
\end{equation} 
where we write $n=n_r+in_i$ where $n_r$ and $n_i$ are the real part and the imaginary part of $n$, respectively.
It leads to
\begin{equation}
k_{iz}\longrightarrow k_{iz}^*,
\end{equation}
i.e., a complex conjugation.

Therefore, under a spatial inversion, we can observe that $M^{(s)}_{tot}\rightarrow M_{tot}^{(s)*}$.
We state that 
 \begin{equation}
 \begin{split}
 M^{(s)*}&\sigma M^{(s)} =\sigma.
 \end{split}
 \end{equation}
For TM polarization, the corresponding result can be obtained in the same manner as $ M^{(p)*}\sigma M^{(p)}=\sigma$.\\
  
 Accordingly, we can derive
 \begin{equation}
 \begin{split}
 \begin{cases}
M_{12}^{(s,p)*}&=-M^{(s,p)}_{21}\\
M_{22}^{(s,p)*}&=M^{(s,p)}_{22}\\
M_{11}^{(s,p)*}&=M^{(s,p)}_{11}\\
\end{cases}
 \end{split}
 \end{equation}

 \section{Appendix B}
We consider an electromagnetic plane wave with TE (i.e., \textbf{s}) polarization propagating in a medium characterized by $\epsilon_1$ and $\mu_1$. The $y$ component of the electric field and the $x$ component of the magnetic field are given by
\begin{equation}
\begin{split}
 \begin{cases}
E^{(s)}_y(x,z)&=E_0^{+}e^{ik_1\cos\theta_1 z +ik_1\sin\theta_1 x}+E_0^{-}e^{-ik_1\cos\theta_1 z +ik_1\sin\theta_1 x}\\
H^{(s)}_x(x,z)&=-\frac{\cos\theta_1}{\eta_1}\{ E_0^{+}e^{ik_1\cos\theta_1 z +ik_1\sin\theta_1 x}\\
&-E_0^{-}e^{-ik_1\cos\theta_1 z +ik_1\sin\theta_1 x}\}\\
\end{cases}
\end{split}
\end{equation}
where  $\eta_1=\sqrt{\frac{\mu_1}{\epsilon_1}}$ is intrinsic impedance of the material, $\theta_1$ is wave propagation direction with respect to z-axis,
 $E_0^{+}$ and $E_0^{-}$ are complex amplitudes of the forward ($+z$) and backward ($-z$) propagating waves, respectively.
Given that electric field,  the magnetic field can be derived by Faraday's law.

Based on the continuity of the electric and magnetic fields established along an interface, we define a wave impedance for TE polarization at any z  by
\begin{equation}
Z^{(s)}(z)\equiv-\frac{E^{(s)}_y(x,z)}{H^{(s)}_x(x,z)}.
\end{equation}
Next, we derive the propagation matrix.
We thus rewrite the total fields at $z$,
\begin{equation}\label{impedancematrix}
\begin{split}
\begin{cases}
E^{(s)}_y(z)&=E^{+}(z)+E^{-}(z)\\
H^{(s)}_x(z)&=-\frac{\cos\theta_1}{\eta_1}(E^{+}(z)-E^{-}(z))
\end{cases}
\end{split}
\end{equation}
where $E^{+}(z)=E_0^{+}e^{i(k_1\cos\theta_1z+k_1\sin\theta_1 x)}$ and $E^{-}(z)=E_0^{-}e^{i(-k_1\cos\theta_1 z+k_1\sin\theta_1 x)}$.
 
With Eq.(\ref{impedancematrix}), we obtain the propagation matrix  as follows
\begin{widetext}
\begin{equation}\label{propagation}
\begin{split}
\begin{bmatrix}
E^{(s)}_y(z)\\
H^{(s)}_x(z)
\end{bmatrix}&=\begin{bmatrix}
1 & 1\\
-\frac{\cos\theta_1}{\eta_1} &\frac{\cos\theta_1}{\eta_1}
\end{bmatrix}\begin{bmatrix}
E^{+}(z)\\
E^{-}(z)
\end{bmatrix}\\
&=\begin{bmatrix}
1 & 1\\
-\frac{\cos\theta_1}{\eta_1} &\frac{\cos\theta_1}{\eta_1}
\end{bmatrix}\begin{bmatrix}
e^{ik_1\cos\theta_1 z} & 0\\
0 & e^{-ik_1\cos\theta_1 z}
\end{bmatrix}
\begin{bmatrix}
E^{+}(0)\\
E^{-}(0)
\end{bmatrix}\\
&=\begin{bmatrix}
1 & 1\\
-\frac{\cos\theta_1}{\eta_1} &\frac{\cos\theta_1}{\eta_1}
\end{bmatrix}\begin{bmatrix}
e^{ik_1\cos\theta_1 z} & 0\\
0 & e^{-ik_1\cos\theta_1 z}
\end{bmatrix}\begin{bmatrix}
1 & 1\\
-\frac{\cos\theta_1}{\eta_1} &\frac{\cos\theta_1}{\eta_1}
\end{bmatrix}^{-1}\begin{bmatrix}
E^{(s)}_y(0)\\
H^{(s)}_x(0)
\end{bmatrix}\\
&=\begin{bmatrix}
\cos[k_1\cos\theta_1 z] & -\frac{i\eta_1\sin[k_1\cos\theta_1 z]}{\cos\theta_1}\\
-\frac{i\cos\theta_1}{\eta_1}\sin[k_1\cos\theta_1 z] &\cos[k_1\cos\theta_1 z] 
\end{bmatrix}\begin{bmatrix}
E^{(s)}_y(0)\\
H^{(s)}_x(0)
\end{bmatrix}
\end{split}
\end{equation}
\end{widetext}
where we already consider a phase accumulation due to wave propagation, namely, $E^{+}(z)=e^{ik_1\cos\theta_1 z}E^{+}(0)$ and $E^{-}(z)=e^{-ik_1\cos\theta_1 z}E^{-}(0)$.

For the bilayer system illustrated in Fig. 1 (a), with  Eq. (\ref{propagation}), we  derive the corresponding propagation matrix $M^{(WI,s)}$ for TE polarization as follows,
 \begin{widetext}
  \begin{equation}\label{TE}
 \begin{split}
 \begin{bmatrix}
 E^{(s)}_{y}(l)\\
 H^{(s)}_{x}(l)
 \end{bmatrix}&=\begin{bmatrix}
 \cos[k_{2z} l] & -\frac{ik_2\eta_2}{k_{2z}}\sin[k_{2z} l]\\
 -\frac{ik_{2z}}{k_2\eta_2}\sin[k_{2z} l] &\cos[k_{2z} l]
 \end{bmatrix}\begin{bmatrix}
 \cos[k_{1z} l] & -\frac{ik_1\eta_1}{k_{1z}}\sin[k_{1z} l]\\
 -\frac{ik_{1z}}{k_1\eta_1}\sin[k_{1z} l] &\cos[k_{1z} l]
 \end{bmatrix}\begin{bmatrix}
 E^{(s)}_{y}(-l)\\
 H^{(s)}_{x}(-l)
 \end{bmatrix}\\
 &=M^{(WI,s)}\begin{bmatrix}
 E^{(s)}_{y}(-l)\\
 H^{(s)}_{x}(-l)
 \end{bmatrix}
 \end{split}
 \end{equation}
 \end{widetext}
where $\eta_i=\sqrt{\frac{\mu_i}{\epsilon_i}}$ is intrinsic impedance of the medium in the i-th slab.
 
 Applying the APT condition, we find 
 $k_{1z}=
 -k^{*}_{2z}\equiv \nu$.
 Consequently, the propagation matrix $M^{(WI,s)}$ for generic APT bilayer systems is given by
\begin{equation}
\begin{split}
\begin{cases}
M^{(WI,s)}_{11}&=\cos[\nu l]\cos[\nu^{*}l]+\frac{\nu}{\nu^{*}}\sin[\nu l]\sin[\nu^{*}l]\\
M^{(WI,s)}_{12}&=-i\{\frac{\mu \omega}{\nu}\cos[\nu^{*} l]\sin[\nu l]-\frac{\mu \omega}{\nu^{*}}\cos[\nu l]\sin[\nu^{*} l]\}\\
M^{(WI,s)}_{21}&=-i\{\frac{\nu}{\mu \omega}\cos[\nu^{*} l]\sin[\nu l]-\frac{\nu^{*}}{\mu \omega}\cos[\nu l]\sin[\nu^{*} l]\}\\
M^{(WI,s)}_{22}&=\cos[\nu l]\cos[\nu^{*}l]+\frac{\nu^{*}}{\nu}\sin[\nu l]\sin[\nu^{*}l].
\end{cases}
\end{split}
\end{equation} 

For TM polarization incidence, the wave impedance at arbitrary z is defined by 
\begin{equation}
Z^{(p)}(z)\equiv\frac{E^{(p)}_x(x,z)}{H^{(p)}_y(x,z)}
\end{equation}
 here $E^{(p)}_x(x,z)$ is x-component of total electric field and $H^{(p)}_y(x,z)$ is y-component of total magnetic field.
 
By invoking the electromagnetic duality theorem and the APT condition, the propagation matrix for a generic APT bilayer system under TM polarization is obtained as
\begin{equation}
\begin{split}
\begin{bmatrix}
E^{(p)}_x(l)\\
H^{(p)}_y(l)
\end{bmatrix}
&=
M^{(WI,p)}\begin{bmatrix}
E^{(p)}_y(-l)\\
H^{(p)}_x(-l)
\end{bmatrix}
\end{split}
\end{equation} 
 where 
 \begin{equation}
\begin{split}
\begin{cases}
M^{(WI,p)}_{11}&=\vert\cos[\nu l]\vert^2+\frac{\epsilon\nu^{*}}{\epsilon^{*}\nu}\vert\sin[\nu l]\vert^2\\
M^{(WI,p)}_{12}&=\frac{i\nu}{\omega\epsilon}\sin[\nu l]\cos[\nu^{*} l]-\frac{i\nu^{*}}{\omega\epsilon^{*}}\sin[\nu^{*} l]\cos[\nu l]\\
M^{(WI,p)}_{22}&=\vert\cos[\nu l]\vert^2+\frac{\epsilon^{*}\nu}{\epsilon\nu^{*}}\vert\sin[\nu l]\vert^2\\
M^{(WI,p)}_{21}&=\frac{i\omega\epsilon}{\nu}\sin[\nu l]\cos[\nu^{*} l]-\frac{i\omega\epsilon^{*}}{\nu^{*}}\sin[\nu^{*} l]\cos[\nu l].\\
\end{cases}
\end{split}
 \end{equation}

With above relation, we further derive a relation of the wave impedances at $z=l$ and $z=-l$,
 \begin{equation}\label{boundaryTM}
 Z^{(p)}(l)=\frac{M^{(WI,p)}_{11}Z^{(p)}(-l)+M^{(WI,p)}_{12}}{M^{(WI,p)}_{21}Z^{(p)}(-l)+M^{(WI,p)}_{22}}.
 \end{equation}
 
In case of reflectionless, the corresponding wave impedance conditions at $z=-l$ and $z=l$ are $Z(-l)=\eta_0\cos\theta_{in}$ and $Z(2l)=\eta_0\cos\theta_{in}$, respectively.
Substituting these conditions into Eq.~(\ref{boundaryTM}), we obtain the reflectionless condition for TM polarization,
\begin{widetext}
\begin{equation} 
\begin{split}
2iIm[\frac{\nu \epsilon^{*}}{\nu^{*}\epsilon}]\vert \sin(\nu l)\vert^2
=2 Re[\frac{i\nu}{\eta_0 k_0c\epsilon\cos\theta_{in}}\cos[\nu^{*}l]\sin[\nu l]]+2Re[\frac{-i\epsilon k_0 c\eta_0\cos\theta_{in}}{\nu}\cos[\nu^{*}l]\sin[\nu l].
\end{split}
\end{equation}
\end{widetext}
The above result holds only when $\epsilon$ and $\mu$ are real.

  \section{Appendix C}
Based on Eq.~(\ref{TE}), we formulate the reflection coefficient for left incidence under TE polarization,
 \begin{equation}\label{TEreflection}
 \begin{split}
 r_L^{(s)}=\frac{(M^{(WI,s)}_{22}-M^{(WI,s)}_{11})+(\frac{M^{(WI,s)}_{12}}{Z(l)}-M^{(WI,s)}_{21}Z(l))}{(M^{(WI,s)}_{22}+M^{(WI,s)}_{11})+(\frac{M^{(WI,s)}_{12}}{Z(l)}+M^{(WI,s)}_{21}Z(l))}.
 \end{split}
 \end{equation}
 
For non-Hermitian systems, the relative permittivity can be expressed as
\begin{equation}
\epsilon= \epsilon^{'}+i\epsilon^{''}
\end{equation}
here $\epsilon^{'}$ and $\epsilon^{''}$ are real.
We now consider $\vert \epsilon^{''}\vert<<1$.
With a  small $\epsilon^{''}$ introduced, we can express the z component of wave number by
\begin{equation}
\begin{split}
&k_{1z}\equiv\nu=\sqrt{(k_0^2\mu_r\epsilon_r^{'}-k_{0x}^2)[1+\frac{ik_{0x}^2\mu_r\epsilon_r^{''}}{k_0^2\mu_r\epsilon_r^{'}-k_{0x}^2}]}\\
&\sim \sqrt{k_0^2\mu_r\epsilon_r^{'}-k_{0x}^2}+\frac{i}{2}\frac{k_0^2\mu_r\epsilon_r^{''}}{\sqrt{k_0^2\mu_r\epsilon_r^{'}-k_{0x}^2}}\equiv \nu^{'}+i\nu^{''}
\end{split}
\end{equation}
to the first order.
We note that $\nu^{'}=\sqrt{k_0^2\mu_r\epsilon_r^{'}-k_{0x}^2}$ and $\nu^{''}=\frac{1}{2}\frac{k_0^2\mu_r\epsilon_r^{''}}{\sqrt{k_0^2\mu_r\epsilon_r^{'}-k_{0x}^2}}$ are real.
Noticeably, $\vert \nu^{''}\vert <<1$.

With $\vert \nu^{''}\vert <<1$, we can approximate
\begin{equation}
\begin{split}
\begin{cases}
\cos[\nu l]&\sim\cos[\nu^{'}l]-i\nu^{''}l\sin[\nu^{'}l]\\
\cos[\nu^{*} l]&\sim\cos[\nu^{'}l]+i\nu^{''}l\sin[\nu^{'}l]\\
\sin[\nu l]&\sim\sin[\nu^{'}l]+i\nu^{''}l\cos[\nu^{'}l]\\
\sin[\nu^{*} l]&\sim\sin[\nu^{'}l]-i\nu^{''}l\cos[\nu^{'}l]\\
\frac{1}{\nu}&\sim\frac{1}{\nu^{'}}-i\frac{\nu^{''}}{\nu^{'2}}\\
\frac{1}{\nu^{*}}&\sim\frac{1}{\nu^{'}}+i\frac{\nu^{''}}{\nu^{'2}}\\
\frac{\nu}{\nu^{*}}&\sim 1+2i\frac{\nu^{''}}{\nu^{'}}
\end{cases}
\end{split}
\end{equation} 
to the first order.
 
Consequently, the corresponding propagation matrix can be approximately as
 \begin{equation}
 \begin{split}
 \begin{cases}
 M^{(WI,s)}_{11}&\sim 1+2i\frac{\nu^{''}}{\nu^{'}}\sin^2[\nu^{'}l]\\
 M^{(WI,s)}_{22}&\sim 1-2i\frac{\nu^{''}}{\nu^{'}}\sin^2[\nu^{'}l]\\
 M^{(WI,s)}_{12}&\sim \mu k_0c\frac{\nu^{''}}{\nu^{'2}}(2l\nu^{'}-\sin[2\nu^{'}l])\\
  M^{(WI,s)}_{21}&\sim \frac{\nu^{''}}{\mu k_0c}(2l\nu^{'}+\sin[2\nu^{'}l]).\\
 \end{cases}
 \end{split}
\end{equation}  

Substituting these into Eq.(\ref{TEreflection}),  the reflection coefficient for left incidence is given by
\begin{widetext}
\begin{equation}
r_L^{(s)}\approx -2i\frac{\nu^{''}}{\nu^{'}}\sin^2[\nu^{'}l]+\cos\theta_{in}\mu_r k_0\frac{\nu^{''}}{2\nu^{'2}}(2\nu^{'}l-\sin[2\nu^{'}l])-\frac{\nu^{''}}{2\cos\theta_{in}\mu_r k_0}(2\nu^{'}l+\sin[2\nu^{'}l]).
\end{equation}
\end{widetext}

In TM polarization,  the corresponding reflection coefficient for left incidence is
\begin{equation}
r_L^{(p)}=\frac{(M^{(WI,p)}_{22}-M^{(WI,p)}_{11})+(M^{(WI,p)}_{21}Z_0-\frac{M^{(WI,p)}_{12}}{Z_0})}{-(M_{22}+M^{(WI,p)}_{11})+(\frac{M^{(WI,p)}_{12}}{Z_0}+Z_0M_{21})}.
\end{equation}

In the presence of a small imaginary part $\lvert \epsilon'' \rvert \ll 1$, the propagation matrix for TM polarization, i.e., Eq.(29), can be approximated as
 \begin{equation}
 \begin{split}
 \begin{cases}
M^{(WI,p)}_{11}\sim 1+2i\sin^2[\nu^{'} l](\frac{\epsilon^{''}}{\epsilon^{'}}-\frac{\nu^{''}}{\nu^{'}})\\
M^{(WI,p)}_{22}\sim 1-2i\sin^2[\nu^{'} l](\frac{\epsilon^{''}}{\epsilon^{'}}-\frac{\nu^{''}}{\nu^{'}})\\
M^{(WI,p)}_{12}\sim -\frac{1}{k_0c\epsilon^{'2}}[2l\epsilon^{'}\nu^{'}\nu^{''}-\sin[2\nu^{'}l](\nu^{'}\epsilon^{''}-\epsilon^{'}\nu^{''})]\\
M^{(WI,p)}_{21}\sim \frac{k_0c}{\nu^{'2}}[-2l\epsilon^{'}\nu^{'}\nu^{''}-\sin[2\nu^{'}l](\nu^{'}\epsilon^{''}-\epsilon^{'}\nu^{''})].\\
 \end{cases}
 \end{split}
 \end{equation}

Consequently, the reflection coefficient for  left incidence is 
\begin{widetext}
\begin{equation}
\begin{split}
r_L^{(p)}
&\approx 2i\sin^2[\nu^{'}l](\frac{\epsilon_r^{''}}{\epsilon_r^{'}}-\frac{\nu^{''}}{\nu^{'}})-\frac{k_0\cos\theta_{in}}{2\nu^{'2}}[-2l\epsilon_r^{'}\nu^{'}\nu^{''}-\sin[2\nu^{'}l](\nu^{'}\epsilon_r^{''}-\epsilon_r^{'}\nu^{''})]-\frac{1}{2k_0\cos\theta_{in}\epsilon_r^{'}}[2l\nu^{'}\nu^{''}-\sin[2\nu^{'}l](\nu^{'}\frac{\epsilon_r^{''}}{\epsilon_r^{'}}-\nu^{''})].
\end{split}
\end{equation}
\end{widetext}

 \end{document}